\documentclass[useAMS,usenatbib]{mn2e}
\usepackage{graphicx}
\usepackage{float}
\usepackage[usenames,dvipsnames]{xcolor}

\title[IRAS~13224-3809 flux-dependent reverberation lags]{Revealing the X-ray source in IRAS~13224-3809 through flux-dependent reverberation lags}
\author[Kara et al.]{E. Kara$^{1}$\thanks{E-mail:
ekara@ast.cam.ac.uk}, A. C. Fabian$^{1}$, E. M. Cackett$^{2}$, G. Miniutti$^{3}$ and P. Uttley$^{4}$\\
$^{1}$Institute of Astronomy, Madingley Rd, Cambridge CB3 0HA\\
$^{2}$Department of Physics and Astronomy, Wayne State University, Detroit, MI 48201, USA\\
$^{3}$Centro de Astrobiologia (CSIC-INTA), Dep. de Astrofisica; LAEFF, PO Box 78, E-28691, Villanueva de la Ca{\~n}ada, Madrid, Spain\\
$^{4}$Astronmical Institute `Anton Pannekoek', University of Amsterdam, Postbus 94249, 1090 GE Amsterdam, the Netherlands}
\begin{document}

\date{Accepted 2012 December 29.  Received 2012 December 13; in original form 2012 October 18}

\pagerange{\pageref{firstpage}--\pageref{lastpage}} \pubyear{2012}

\maketitle

\label{firstpage}

\begin{abstract}
IRAS~13224-3809 was observed in 2011 for 500~ks with the {\em XMM-Newton} observatory.  We detect highly significant X-ray lags between soft (0.3 -- 1 keV) and hard (1.2 -- 5 keV) energies.  The hard band lags the soft at low frequencies (i.e. hard lag), while the opposite (i.e. soft lag) is observed at high frequencies.  In this paper, we study the lag during flaring and quiescent periods. We find that the frequency and absolute amplitude of the soft lag is different during high-flux and low-flux periods.  During the low flux intervals, the soft lag is detected at higher frequencies and with smaller amplitude.  Assuming that the soft lag is associated with the light travel time between primary and reprocessed emission, this behaviour suggests that the X-ray source is more compact during low-flux intervals, and irradiates smaller radii of the accretion disc (likely because of light bending effects).  We continue with an investigation of the lag dependence on energy, and find that isolating the low-flux periods reveals a strong lag signature at the Fe K$\alpha$ line energy, similar to results found using 1.3 Ms of data on another well known Narrow-Line Seyfert I galaxy, 1H0707-495.

\end{abstract}


\section{Introduction}
\label{intro}

The X-ray emitting region in accreting black hole systems is not well understood.  Generally, it is agreed that the angular momentum of accreting gas onto a black hole forms an optically thick, geometrically thin accretion disc that radiates as a series of blackbody components \citep{shakura73}. For supermassive black holes, this thermal emission peaks in the UV.  Some of the thermal disc photons are then inverse Compton upscattered to X-ray energies by mildly relativistic electrons in a hot cloud, called the corona \citep{haardt91,haardt93}.  While the X-ray emission mechanism is relatively well understood, the geometry of the corona where the X-rays are produced is an area of active research.  In this paper, we probe the geometry of the corona through an analysis of the X-ray reverberation lags. 

X-ray variability studies have been influential in starting to understand the size of the emitting region. AGN are known to be extremely variable in the X-ray band, commonly doubling in amplitude on timescales of less than one day for radio-quiet sources \citep{mchardy88}.  Substantial variability occurs on timescales of the light crossing time of the source (or greater), and therefore we can calculate an upper limit on the size of the emitting region, $R < c\delta t$ \citep{terrell67}.  From observations of fast variability in AGN, we know that the corona must be compact, likely within $\sim 100~r_\mathrm{g}$.  Recent observations from the X-ray variability seen from microlensed quasars also show that the corona is compact to within $10~r_\mathrm{g}$ or less \citep{chartas09,dai10,chartas12}.

Variability studies of black hole binaries led to the discovery of the X-ray lag, where usually hard photons lag behind soft photons \citep[e.g.][]{miyamoto88,nowak96,nowak99}. This effect was also later observed in AGN \citep{papadakis01,mchardy04}.  The origin of the hard lag is still not well understood, though in the prevailing model, the multiplicative propagation effects of fluctuations in the accretion flow modulate regions emitting soft photons before those emitting hard photons \citep{kotov01, arevalo06}.

Recent developments in X-ray variability have revealed a new type of lag, where soft photons lag behind hard photons.  This lag, interpreted as a reverberation lag, offers a new perspective in which to study the X-ray emitting region in AGN.  Understanding the phase lag between light curves of different energy bands gives us an orthogonal approach to testing physical models that are often degenerate in the time-integrated energy spectrum. 

Reverberation lags were first discovered in Narrow-Line Seyfert 1 (NLS1) galaxy, 1H0707-495 \citep{fabian09}. Since then, lags have been detected more than a dozen other Seyfert galaxies \citep{emmanoulopoulos11,demarco11,demarco12,zoghbi12}, including most recently IRAS~13224-3809 \citep{fabian12}.  Reverberation lags have also been observed on the order of milliseconds or less in black hole binary GX 339-4 \citep{uttley11}.

The soft lags can be understood with the standard reflection scenario \citep{guilbert88,lightman88}.  In this model, the bulk of the X-ray continuum is produced in the corona.
Some continuum photons fall towards the surface of the accretion disc, are absorbed by photoelectric absorption, and reprocessed into an X-ray reflection spectrum comprised of a reflection continuum and emission lines \citep[e.g. ][]{ross05}.  When reflection arises in the accretion disc, the whole reflection spectrum is blurred due to Doppler and other relativistic effects close to the black hole.  Soft lags, where the direct continuum variations lead those in the soft-band reflection, are then interpreted as the additional light travel time taken by the reflected paths between the corona and the inner accretion disc.  

IRAS~13224-3809 ($z=0.066$) is one of the most X-ray variable Seyfert 1 galaxies known \citep{boller96}, and therefore is a useful source with which to probe the environments of the innermost regions.
It was first observed with {\em XMM-Newton} in 2002 for 64~ks \citep{boller03,gallo04,ponti10}, 
and most recently, for 500~ks, which led to a significant detection of the soft lag \citep{fabian12}. 
In this letter, we follow up that recent work with an analysis of the soft lags of IRAS~13224-3809 in the low and high flux states.

\section{Observations and Data Reduction}
\label{obs}

The {\em XMM-Newton} satellite \citep{jansen01} observed IRAS~13224-3809 for 500~ks over four orbits from 2011 July 19 to 2011 July 29 (Obs. IDs 0673580101, 0673580201, 0673580301, 0673580401).  
For this analysis of the reverberation lags, we focus on the high time-resolution data from the EPIC-pn camera \citep{struder01}.  
The first observation was taken in full window imaging model, and next three in large window imaging mode. 
All of the data were reduced in the same way, using the {\em XMM-Newton} Science Analysis System (SAS v.11.0.0) and the newest calibration files.  
The details of the data reduction are explained in \citet{fabian12}.  

The data were cleaned for high background flares, which resulted in a total exposure time of $\sim$ 300 ks.  The data were selected with the condition {\sc pattern} $\le 4$. Pile-up effects were not significant in any of the observations.

The source light curve was extracted from circular regions of radius 35 arcsec, which were centered on the maximum source emission. The background light curves were chosen from a circular region of the same size, and were the same distance to the readout node as the source region.  The background subtracted light curves were produced using the tool {\sc epiclccorr}.  The light curves ranged in length from 83~ks to 124~ks with 10~s bins. 

\section{Results}
\label{results}

\subsection{Lag vs. Frequency}
\label{lagfreq_sec}

\begin{figure}
\begin{center}
\includegraphics[width=7.5cm]{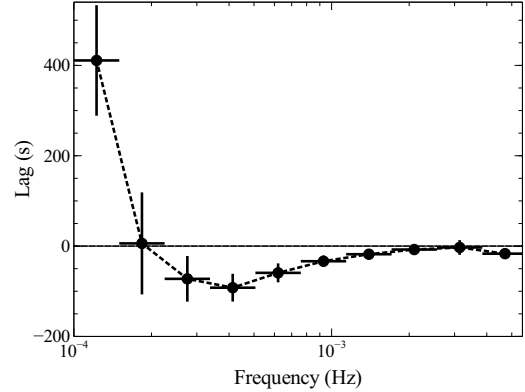}
\caption{Lag-frequency spectrum for the 500 ks observation. The lag is calculated between the soft energy band (0.3 - 1.~keV) and the hard band (1.2 - 5.~keV). 
The most negative lag (at $\nu=4.1 \times 10^{-4}$~Hz) is $-92 \pm 31$ s. }
\label{lag_freq}
\end{center}
\end{figure}

The lag is computed in the usual way following the formulae described in \citet{nowak99}.  The frequency-dependent lag is calculated between light curves in the soft (0.3 -- 1. keV) and hard (1.2 -- 5 keV) energy bands, which are dominated by the soft excess (likely disc-reflection) and the primary powerlaw-like emission, respectively \citep{fabian12,ponti10}.  
The Fourier transform of a light curve contains an amplitude and complex phase, such that the Fourier transform of soft light curve $s(t)$ is written $S=| S | e^{i\phi_s}$.  The cross product of the soft and hard band Fourier transforms, $ H^{\ast} S = | H | | S | e^{i(\phi_s-\phi_h)}$, provides the phase difference between the two energy bands.  This phase difference is converted back into a frequency-dependent time lag, where $\tau(f) = (\phi_s-\phi_h)/2\pi f$.  Given this definition for the lag, a negative amplitude lag means that the soft band lags behind the hard band, while the positive amplitude lag shows the hard band lagging.  We take note that the phase lag is defined over the interval $(-\pi, \pi)$, which causes phase wrapping at high frequencies \citep{nowak96}.

Fig.~\ref{lag_freq} shows frequency-dependent time lag between the soft and hard bands, from \citet{fabian12}.  At low frequencies, we find a large positive lag where the hard flux lags behind the soft flux by hundreds of seconds.  At higher frequencies, $\nu \sim [2 - 10] \times 10^{-4}$ Hz, there is a negative lag where the soft flux lags the hard by $\sim 100$~s. Above $\sim 1.5 \times 10^{-3}$ Hz, the signal becomes dominated by Poisson noise.  

\subsection{Flux-resolved analysis}
\label{flare_sec}

\begin{figure*}
\includegraphics[width=\textwidth]{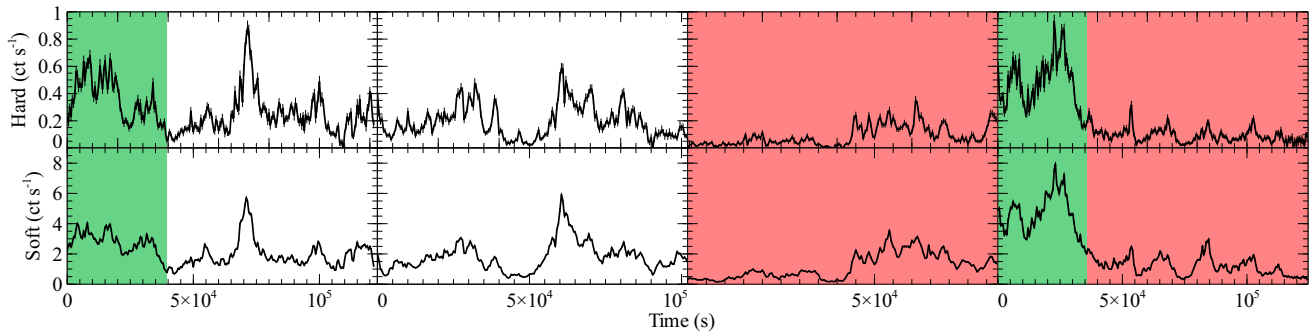}
\caption{Hard band (1.2 -- 5 keV) and soft band (0.3 -- 1 keV) light curves for the 4 orbits in 500~s bins.  Green regions denote high flux segments and red regions denote low flux segments.  These high and low flux segments will be used for the lag analysis shown in Fig.~\ref{lag_flare}.}
\label{lc}
\end{figure*}

IRAS~13224-3809 is a highly variable source that exhibited distinct flaring and quiescent periods during the 500~ks observation.  In this section, we examine the lags from low and high flux intervals.  Fig.~\ref{lc} shows light curves from the 4 orbits.  The low flux sample is shown in red, and the high flux is shown in green.  We chose light curve segments that were long in order to obtain information at low frequencies. 
The power spectral density confirms that up to $1.5 \times 10^{-3}$ Hz, we are not in a Poisson noise dominated frequency regime for both the low and high flux samples.

\subsubsection{The lag-frequency spectrum}
\label{fluxlag_sec}

\begin{figure}
\begin{center}
\includegraphics[width=7.5cm]{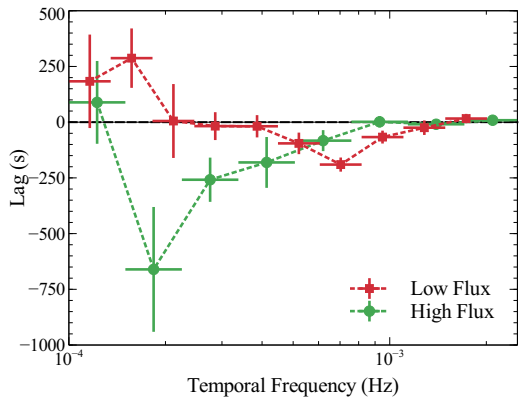}
\caption{Lag-frequency spectrum for the low (red) and high (green) flux segments shown in Fig.~\ref{lc}. The most negative lag for the high flux intervals is $-660 \pm 280$~s at $\nu=1.8 \times 10^{-4}$ Hz, and the most negative lag for the low flux is higher: $-230 \pm 31$~s at $\nu=7 \times 10^{-4}$ Hz. Again, note that the most negative lag for the total observation (Fig.~\ref{lag_freq}) occurs at $\nu = 4.1 \times 10^{-4}$ Hz. }
\label{lag_flare}
\end{center}
\end{figure}

\begin{figure}
\begin{center}
\includegraphics[width=7.5cm]{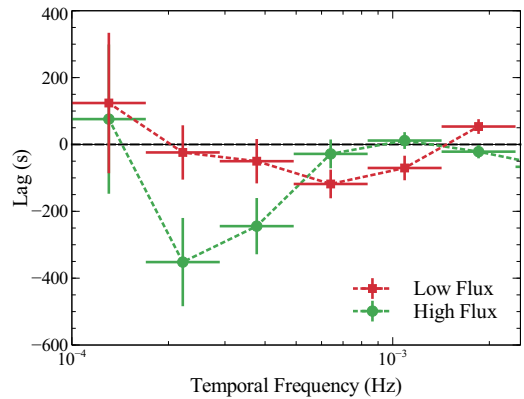}
\caption{Lag-frequency spectrum of the 4th orbit alone, for the low (red) and high (green) flux segments shown in Fig.~\ref{lc}.  The most negative lag for the high flux interval is $-350 \pm 130$~s at $\nu=2.2 \times 10^{-4}$ Hz, and the most negative lag for the low flux is higher: $-118 \pm 42$~s at $\nu=6.5 \times 10^{-4}$ Hz.}
\label{obs4}
\end{center}
\end{figure}

Fig.~\ref{lag_flare} shows the lag vs. frequency for the high and low flux segments.  The lag is computed in the same way as in Section~\ref{lagfreq_sec}, between the energy bands of 0.3 -- 1 keV and 1.2 -- 5 keV.  Focusing on the high flux points in green, we see that the soft lag occurs at lower frequencies than found from the total light curve sample, in Fig.~\ref{lag_freq}.  This means that the variability associated with the soft lag occurs on longer timescales during the flaring period.  Also, the amplitude of the lag is greater.
Now looking at the red low flux points, we see that the soft lag occurs at higher frequencies than the total, indicating the source variability is occurring on shorter timescales.

As a check, we compute flux-dependent lags using orbit 4 alone, which shows distinct low and high flux segments (as designated in green and red in the rightmost panel of Fig.~\ref{lc}).  The lag is computed for these two sets of continuous light curves, and shown in Fig.~\ref{obs4}.  We find that for continuous segments (i.e. for one realisation of some underlying process), the same flux-dependent behaviour of the lag is clear, just with lower signal-to-noise. The analysis that follows has been performed with low and high flux segments from the total 500~ks observation.

\subsubsection{The lag-energy spectrum}
\label{lagen_sec}

\begin{figure*}
\includegraphics[width=\textwidth]{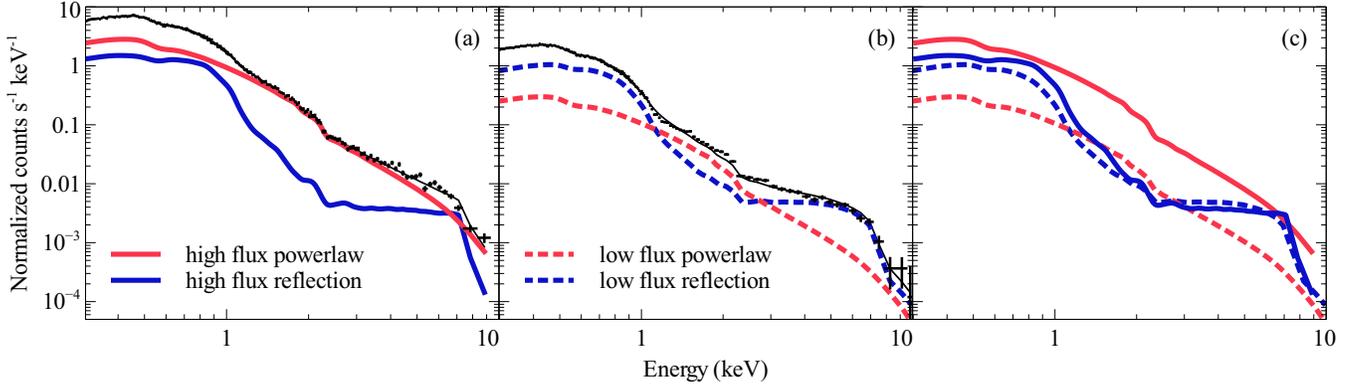}
\caption{Energy spectra for the high and low flux samples in panels (a) and (b) respectively.  For clarity, only the power-law and reflection components are plotted.  Panel (c) shows the powerlaw and reflection for the low and high flux spectra over plotted.} 
\label{spec}
\end{figure*}

\begin{figure} 
\begin{center}
\includegraphics[width=7.5cm]{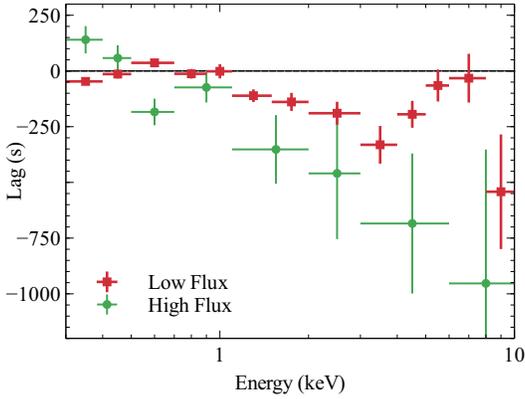}
\caption{Lag-energy spectra for the low-flux (red) and high-flux (green) intervals, showing the energy dependence of the lag for the frequencies of the soft lag ($\nu=[5.8 - 10.5] \times 10^{-4}$ Hz for the low flux, and $\nu=[1.4 - 2.8] \times 10^{-4}$ Hz for the high flux). }
\label{lag_en}
\end{center}
\end{figure}

We also examine how the flux-dependent lag evolves with energy at the frequencies of the negative lag (i.e. $[5.8-10.5] \times 10^{-4}$ Hz for the low flux, and $[1.4-2.8] \times 10^{-4}$ Hz for high fluxes). In this analysis, we measure the time lag of light curves in relatively narrow energy bins with respect to the light curve of a broader reference band from 0.3 -- 0.8~keV \citep[see][ for further details]{kara12}. We use the convention that a positive lag means that the light curve in that energy bin lags behind the reference band, while a negative lag means that the light curve in that energy bin leads the reference.

Fig.~\ref{lag_en} shows the high-frequency lag-energy spectrum for the low and high flux segments.  The low flux segments show a much clearer signal than the high flux, likely due to poor statistics and lower coherence in the high flux.  The general trends of the two lag-energy spectra are similar, except that the amplitude of the lag between 1--4 keV is greater for the high flux.  The low-flux lag-energy spectrum shows a clear peak at $\sim~6.5$~keV, the energy of the Fe K$\alpha$ line. There does not appear to be a corresponding peak at high fluxes, but as the error bars are so large, it cannot be ruled out.

\subsubsection{The time-integrated energy spectrum}

We complement the lag analysis presented above with a brief investigation of the spectral properties in the low and high flux segments.  We applied the best fit spectral model from \citet{fabian12} to the low and high flux spectra, freezing all physical parameters that are not likely to change between state (i.e. inclination, galactic absorption, Fe abundance).  The results are shown in panels (a) and (b) of Fig.~\ref{spec} for the high and low flux segments, respectively.  For clarity, only the power-law and total reflection components are plotted.  The high flux spectrum is dominated by the powerlaw, while the low flux spectrum, shows a greater contribution from reflection.  

Panel (c) of Fig.~\ref{spec} shows the power-law and reflection components of the low and high flux segments overplotted. We notice that the power-law flux increases by an order of magnitude between the low and high flux spectra, but the reflection component remains nearly unchanged, especially at the energy of the Fe~K line.

In \citet{kara12}, we use Monte Carlo simulations to show that dilution will have an effect on the measurement of the lag.  The lag is measured between a soft energy band and a hard band, where, to first order, we approximate the soft band as reflection and the hard band as powerlaw. However, we understand that there is `contamination' in each band caused by the other varying component.  As these components have different intrinsic time delays, the net measured lag will be some weighted average of all the components.  In the ideal case, where there is no contamination, the intrinsic lag will be the measured lag. In the other extreme, where there are equal parts of reflection and powerlaw in each band, the measured lag will be zero.  Furthermore, contamination by a non-varying or an uncorrelated component will not effect the measurement of the lag, as the lag is measured between coherent signals.

It is clear from the spectra in Fig.~\ref{spec} that we need to account for dilution in the measurement of the lag.  Here we quantify the amount of dilution by measuring the reflection fraction in the hard and soft bands. For example, in the low flux spectrum, the soft band is composed of 80\% reflection, but also the hard band is composed of 45\% reflection.  The net amount of reflection between the two bands is therefore 35\%, meaning that the measured lag is only 35\% of the intrinsic lag. Similarly for the high flux spectrum, the soft band is composed of 40\% reflection, while the hard band is contaminated by 10\% reflection. Again the net amount of reflection is 30\%, meaning that the measured lag in the high flux is only 30\% of the intrinsic lag.  We conclude then that the effect of dilution is nearly the same for both the high and low flux lags, and therefore we can be confident that the intrinsic amplitude of the lag in the high flux sample is indeed greater than the lag in the low flux.

\section{Discussions}
\label{discuss}

\begin{figure}
\begin{center}
\includegraphics[width=7.5cm]{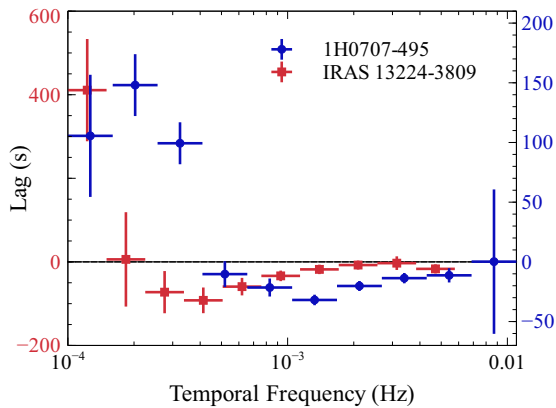}
\caption{The lag-frequency spectrum for the total observation (same as Fig.~\ref{lag_freq}) in red, overplotted with the lag-frequency spectrum of 1H0707-495 from \citet{kara12} in blue. }
\label{lag_freq_1h}
\end{center}
\end{figure}

\begin{enumerate}
\item As shown in \citet{fabian12}, IRAS~13224-3809 exhibits frequency dependent lags where the hard band lags the soft band at low frequencies below $\sim 2 \times 10^{-4}$~Hz, while the soft band lags the hard band at higher frequencies (Fig.~\ref{lag_freq}).
\item In this work, we take a step further and study the flux-dependence of the observed lags. We find that the soft lag has a larger amplitude and occurs on longer timescales during high flux states than during low flux states (Fig.~\ref{lag_flare}).
\item According to the lag interpretation presented by \citet{fabian09}, \citet{zoghbi10} and others, soft lags are reverberation lags due to the light travel time between the primary source and the inner accretion disc.  Therefore, the low flux state (smaller reverberation lags at shorter timescales) refers to a compact emitting source that is closer to the central black hole, while the high flux state (bigger lag at longer timescales) refers to an extended emitting region further from the black hole.
Moreover, we point out that a more compact corona will inevitably irradiate smaller disc radii because of light bending \citep[e.g.][]{miniutti04}, thus providing a qualitatively self-consistent explanation for both the higher frequency and smaller amplitude of the soft lags during low flux states.
\end{enumerate}

The spatial extent of the corona can be interpreted as an increase in the height of the corona above the disc, or some radial expansion of the corona, or some combination of the two.  An increase in the amplitude of the lag is easily understood for a vertically extended corona as the light travel between the corona and the disc increases. For a radially extended source, one can imagine that the corona irradiates a larger portion of the disc, causing a longer average lag.  Some complexities may arise with this interpretation because the emissivity decreases greatly with larger disc radius, however we cannot rule out a radially extended corona from the observations presented here.  Both the lag amplitude and the frequency change by a factor of $\sim 3.5$ between low and high fluxes, which could imply that the corona extends 3.5 times further away in the high state than in the low.  

The energy spectra of the low and high flux segments (Fig.~\ref{spec}) agree with the interpretation of an extended corona.  The low flux sample shows a stronger contribution from reflection. This is expected from a more compact corona, where light bending effects close to the central black hole cause a greater fraction of continuum emission to be directed towards the disc. Spectral modelling of 1H0707-495 during a low flux state in 2011 also shows a more compact corona \citep{fabian12a}.  

The spectral difference between low and high flux samples appears to be caused by a change in powerlaw flux, and not from a change in reflection.  Long timescale variations in continuum that are decoupled from reflection have also been observed in MCG-6-30-15 \citep{vaughan04,miniutti07}, and explained within the framework of strong light bending.  
We note that while the reflection spectrum appears unchanged on these long timescales, the frequency-resolved covariance spectrum \citep[as shown for 1H0707-495 in][]{kara12} does show that reflection varies with the continuum on short timescales.

\subsection{Comparison with 1H0707-495}

\begin{figure}
\begin{center}
\includegraphics[width=7.5cm]{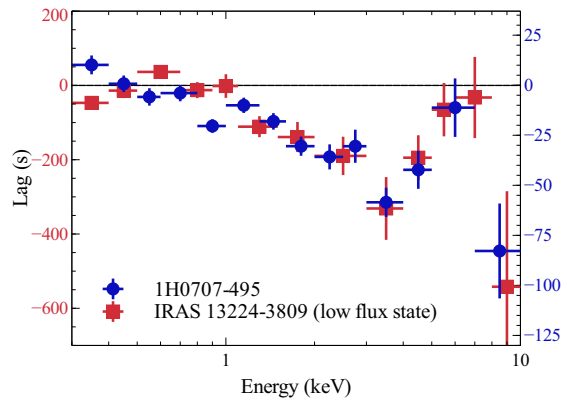}
\caption{High-frequency lag-energy spectrum for the low flux segments in IRAS 13224-3809 (red), overplotted with the high-frequency lag-energy spectrum of 1H0707-495 ($\nu=[0.98 - 2.98] \times 10^{-3}$ Hz) in blue. }
\label{1h}
\end{center}
\end{figure}

The lag analysis of IRAS~13224-3809 shows striking similarities to the results found for 1H0707-495 using 1.3~Ms of data \citep{kara12}.  Fig.~\ref{lag_freq_1h} shows the lag-frequency spectrum of IRAS 13224-3809 overplotted with 1H0707-495 (note different y-axis scales).  We find that IRAS 13224-3809 shows same general positive-to-negative trend as seen in 1H0707-495, just with a larger amplitude negative lag that occurs at lower frequencies.  The negative lag in 1H0707-495 is measured to be $-31.9 \pm 4.2$~s at $1.33 \times 10^{-3}$~Hz, while the negative lag in IRAS~13224-3809 is a factor of $\sim 3.33$ times more ($-92 \pm 31$~s) at a frequency that is $\sim 3.25$ times lower ($4.1 \times 10^{-4}$~Hz).  Both the lag and the frequency change roughly by a factor of 3.3. Assuming a simple mass scaling relationship, this implies that IRAS~13224-3809 is 3.3 times more massive that 1H0707-495. According to the mass scaling trends in \citet{demarco12}, the lags in IRAS~13224-3809 are consistent with a black hole mass of $10^{7}$ M$_{\sun}$. In this work, however, we use IRAS~13224-3809 to show that there is some flux-dependence to the reverberation lag.  Therefore, we should take some caution in predicting the mass of the central black hole from the amplitude and frequency of the lag.  While observations used to measure reverberation lags are relatively long, it is possible to sample some intrinsically high or low flux state. This would introduce systematics to the simple mass scaling relationship.

We compare the low-flux lag-energy spectrum of IRAS~13224-3809 with the lag-energy spectrum of 1H0707-495 (Fig.~\ref{1h}).  The shapes of the spectra for these two different sources are strikingly similar, but the amplitude of the lags is greater for the low flux segments in IRAS~13224-3809.  The peak at $\sim~6$~keV and the dip at 3 -- 4 keV are clearly present in both data sets.  Being able to isolate the low-flux segments of IRAS~13224-3809 has allowed us to clearly uncover the reverberation signal, as was only possible with 1.3 Ms of data with 1H0707-495.  1H0707-495 does not have such distinct low-flux segments as IRAS~13224-3809, and so a similar flux-resolved analysis was not possible for this source. 

The high-frequency lag-energy spectra of IRAS~13224-3809 and 1H0707-495 (Figs.~\ref{lag_en} and \ref{1h}) are consistent with a reflection model, where the negative lags are associated with the light travel time between the continuum-emitting corona and the accretion disc.  The lag-energy spectrum shows the 1 -- 4 keV band (powerlaw dominated) leads the 0.3 -- 0.8 keV reference band (where reflection is appreciable).  Similar to 1H0707-495, we see a distinct peak at $\sim 6$ keV, the rest frame energy of the Fe~K line, and a dip at 3 -- 4~keV, which has been interpreted as a signature from the red wing of the Fe~K line, originating from the innermost radius.  This interpretation is discussed further in \citet{kara12}, where we model the high-frequency lag-energy spectrum of 1H0707-495.

\subsection{Conclusions}

Looking at the flux-resolved reverberation lags reveals new observational constraints that can help us distinguish between the physical nature of the low and high flux states.  We learn that flares are likely connected to the expansion of the corona out vertically and possibly radially, too, and that when the source is in quiescence, the corona remains compact and close to the central region.

\section*{Acknowledgements}

This work is based on observations obtained with {\em XMM-Newton}, an ESA science mission with instruments and contributions directly funded by ESA Member States and NASA.  EK thanks the Gates Cambridge Scholarship. ACF thanks the Royal Society.

\label{lastpage}

\end{document}